\documentclass[12pt]{article}
\usepackage{amsfonts}
\advance\oddsidemargin by -1in
\advance\evensidemargin by -1in
\advance\topmargin  by -0.5in
\newcommand{\be}{\begin{equation}}
\newcommand{\ee}{\end{equation}}
\newcommand{\ba}{\begin{array}}
\newcommand{\ea}{\end{array}}
\newcommand{\BB}{{\rm B}}
\newcommand{\CC}{\mathbb{C}}
\newcommand{\RR}{\mathbb{R}}
\newcommand{\calL}{{\cal L }}

\newcommand{\calH}{{\cal H }}
\newcommand{\calS}{{\cal S }}
\newcommand{\la}{\langle}
\newcommand{\ra}{\rangle}

\newcommand{\ep}{\epsilon}
\newcommand{\da}{\downarrow}

\title{Requirements for compatibility between local and
multipartite quantum states} 

\author{Sergey Bravyi\thanks{e-mail: serg@cs.caltech.edu}\\
{\it Institute for Quantum Information,}\\
{\it California Institute of Technology,}\\
{\it Pasadena, CA, 91125, USA} }

\date{\today}

\begin{document}

\maketitle

\begin{abstract}
We consider a partial trace transformation which maps a 
multipartite quantum state to collection of local density
matrices. We call this collection a {\it mean field state}.
For the Hilbert spaces
$(\CC^2)^{\otimes n}$ and $\CC^2\otimes\CC^2\otimes\CC^4$
the necessary and sufficient conditions under which a
mean field state is compatible with at least one
multipartite  pure state are found.
Compatibility of mean field states with more general classes
of multipartite quantum states is discussed.
\end{abstract}

\section{Introduction and main results}

Consider a system consisting of $n$ quantum particles and denote
$\calH^{(i)}$ a Hilbert space of states of the  $i$-th particle.
A state of the whole system can be most exhaustively
described by a vector $|\Psi\ra$ from the Hilbert space
$$
\calH = \calH^{(1)}\otimes\cdots\otimes \calH^{(n)}
$$
(a pure state). However, sometimes we need not such
a detailed description.
For example if we are interested only in expectation
values of one-particle observables, it is sufficient to keep track
of just $n$ reduced density matrices $\rho^{(1)} = tr_{i\ne 1}
(|\Psi\ra\la\Psi|),\ldots,\rho^{(n)} = tr_{i\ne n}
(|\Psi\ra\la\Psi|)$ describing the states of
the individual particles~(here the
subscript $i$ shows the particles to be traced out). This is some
kind of mean field description because we completely ignore all
types of correlations~(both quantum and classical) between the
different particles.

An arbitrary  collection of $n$ density matrices  $(\rho^{(1)},\ldots,\rho^{(n)})$
will be referred to as a {\it mean field state}.
Naturally, some mean field states can not be represented
as partial traces of any pure state.
For example let us take $n=2$ and $\calH^{(1)}=
\calH^{(2)}$. From the Schmidt decomposition we learn
that the density matrices $\rho^{(1)}$ and $\rho^{(2)}$ 
arising from a pure state
must share the same set of the eigenvalues.
If this is not the case,  we can say
that a mean field state $(\rho^{(1)},\rho^{(2)})$ is not compatible with a
pure multipartite state. Conversely, if the eigenvalues of 
$\rho^{(1)}$ and $\rho^{(2)}$ match, we can always find a pure state $|\Psi\ra$
such that $\rho^{(1)} = tr_2 (|\Psi\ra\la\Psi|)$ and 
$\rho^{(2)} = tr_1 (|\Psi\ra\la\Psi|)$.

The main purpose of this paper is to find the necessary and sufficient
conditions for compatibility between
mean field states and multipartite pure states in the case of 
three or more particles where Shmidt decomposition 
can not be applied. 
For any Hilbert space $\calS$ denote $D(\calS)$ a set of all Hermitian,
non-negative operators with unital trace (a set of density matrices).
Now we can formulate our first problem.

\noindent
{\bf Problem 1.} {\it Given a multipartite system
 $\calH = \calH^{(1)}\otimes\cdots\otimes \calH^{(n)}$,
find a set of all mean field states
$(\rho^{(1)},\ldots,\rho^{(n)})$,
$\rho^{(1)} \in D(\calH^{(1)}),\ldots,\rho^{(n)} \in D(\calH^{(n)})$
which are compatible with at least one multipartite pure state
$|\Psi\ra \in \calH $.}

\noindent
Let us denote the set to be found $M_{pure}$.
Its formal definition is
$$
M_{pure} =\{ (\rho^{(1)},\ldots,\rho^{(n)})
 \ : \ \rho^{(i)} = tr_{j\ne i} (|\Psi\ra\la \Psi|),
\ \ |\Psi\ra \in \calH, \ \ \la \Psi|\Psi\ra=1 \}.
$$

\noindent
The compatibility problem may be generalized to the case when a
state of the whole system is a mixed one with a specified set of the
eigenvalues. 
Denote $eig(\rho)$ a string of eigenvalues of an operator
$\rho$ (to avoid ambiguity, let us agree that the eigenvalues
are arranged in the decreasing order). Denote also
$D(\calS,\lambda) = \{ \rho\in D(\calS) \ : \
eig(\rho) = \lambda \}$
a set of density matrices with a specified spectrum $\lambda$.  
Our second problem is following.

\noindent
{\bf Problem 2.} {\it Given a multipartite system
 $\calH = \calH^{(1)}\otimes\cdots\otimes \calH^{(n)}$ and a
string of eigenvalues $\lambda$,
find a set of all mean field states
$(\rho^{(1)},\ldots,\rho^{(n)})$,
$\rho^{(1)} \in D(\calH^{(1)}),\ldots,\rho^{(n)} \in D(\calH^{(n)})$
which are compatible with at least one multipartite mixed state
$\rho \in D(\calH,\lambda)$.}

\noindent
Let us denote the set to be found $M(\lambda)$.
Its formal definition is
$$
\label{mixed}
M(\lambda) = \{ (\rho^{(1)},\ldots,\rho^{(n)})
\ : \ \rho^{(i)} = tr_{j\ne i} (\rho), \ \
\rho \in D(\calH,\lambda) \}.
$$

\noindent
The Problem~2 can be reduced to the Problem~1 by introducing one additional
particle (ancilla)
with a space of states $\calH$. One suffices to find the set $M_{pure}$
for the extended system and then 
find the subset of $M_{pure}$ corresponding to
the fixed eigenvalues of the ancilla.
In that sense the two problems are equivalent.

\vspace{1cm}
The solution of the Problem~1 for an arbitrary $n$ is known only for
the simplest case $\calH^{(i)}=\CC^2$ for all $i$, see~\cite{Sudbery02}.
This is a system of $n$ spins $\frac12$ or, equivalently, of $n$ qubits.
Let $(\rho^{(1)},\ldots,\rho^{(n)})$ be an arbitrary mean field state
of the system. Denote $\lambda^{(i)}$
the lowest eigenvalue of $i$-th qubit's density matrix $\rho^{(i)}$.
By definition, $\lambda^{(i)} \in [0,\frac12]$.  Then
$(\rho^{(1)},\ldots,\rho^{(n)})\in M_{pure}$ if and only if for each $i$
in the range $1,2,\ldots,n$ one has
\be
\label{result1}
\lambda^{(i)} \le
 \sum_{j\ne i} \lambda^{(j)}.
\ee
For instance, if $n=2$, one gets: $\lambda^{(1)}\le \lambda^{(2)}$,
$\lambda^{(2)}\le \lambda^{(1)}$, that is $\lambda^{(1)}=\lambda^{(2)}$
as it should be.
For $n=3$ one gets the triangle inequalities:
$\lambda^{(1)}\le \lambda^{(2)} + \lambda^{(3)}$,
$\lambda^{(2)}\le \lambda^{(3)} + \lambda^{(1)}$,
$\lambda^{(3)}\le \lambda^{(1)} + \lambda^{(2)}$.
For real spin systems, such as magnetics, we have $n\gg 1$
and the inequalities~(\ref{result1}) impose practically no restrictions
on the mean field states. However these inequalities may be important in quantum
information theory, in particular in the study of multipartite
analogs of the Schmidt decomposition~\cite{Carteret,Acin} and
polynomial invariants under local unitary transformations~\cite{Werner,Sudbery00,Barnum}.
In Section~\ref{problem1} we describe an alternative proof of~(\ref{result1})
obtained independently from the work~\cite{Sudbery02}.

We have solved the Problem~2 only for a bipartite system and only in the
simplest case $\calH^{(1)}=\calH^{(2)}=\CC^2$. This is a system of two
qubits. Let us label the qubits by letters A and B. Consider an
arbitrary mean field state $(\rho^A,\rho^B)$. Denote $\lambda^A$
and $\lambda^B$ the lowest eigenvalues of the density matrices
$\rho^A$ and $\rho^B$. By definition, $\lambda^A,\lambda^B \in
[0,\frac12]$. Let $\lambda=\{
\lambda_1,\lambda_2,\lambda_3,\lambda_4 \}$ be an arbitrary string
of real non-negative numbers, such that
$\lambda_1\ge\lambda_2\ge\lambda_3\ge\lambda_4$ and $\sum_{i=1}^4
\lambda_i=1$. In Section~\ref{problem2} we will show that
$(\rho^A,\rho^B)\in M(\lambda)$ if and only if
\be
\label{result2}
\ba{rcl}
\lambda^A &\ge& \lambda_4 + \lambda_3,\\
\lambda^B &\ge& \lambda_4 + \lambda_3,\\
\lambda^A + \lambda^B &\ge& 2\lambda_4 + \lambda_3 + \lambda_2,\\
|\lambda^A - \lambda^B| &\le&
\min \{ \lambda_1 - \lambda_3, \lambda_2 - \lambda_4 \}.\\
\ea
\ee
This result also provides a solution of the Problem~1 for $n=3$ with
$\calH^{(1)}=\calH^{(2)}=\CC^2$ and $\calH^{(3)}=\CC^4$.

Although we can not attack the Problem~2 in the general case, 
we can solve its simplified version, namely find the convex
envelope~\cite{conv} of the set $M(\lambda)$:
$$
Conv(M(\lambda)) = \{ (\rho^{(1)},\ldots,\rho^{(n)})
\ : \ \rho^{(i)} = tr_{j\ne i} (\rho), \ \
\rho \in Conv(D(\calH,\lambda)) \},
$$
where $Conv(D(\calH,\lambda))$ is the convex envelope of the set
$D(\calH,\lambda)$. By definition, any density matrix $\rho\in Conv(D(\calH,\lambda))$
can be represented as  
$\rho=\sum_\alpha p_\alpha U_\alpha \rho_0 U^\dag_\alpha$
for some probabilities $p_\alpha\ge 0$, $\sum_\alpha p_\alpha =1$, unitary
operators $U_\alpha$ and some density matrix $\rho_0\in D(\calH,\lambda)$.
The set $Conv(D(\calH,\lambda))$ is important for example for description
of interaction between the system and the {\it classical} environment.

The simplified Problem~2 is considered in Section~\ref{convex}. We show
that it can be reduced to the analogous {\it classical} problem
concerning multipartite probability distributions. To
describe this reduction, let us consider as an example
a bipartite
system of Alice and Bob, $\calH=\calH^A\otimes \calH^B$.
Let
$\lambda^{A}_j$
be the $j$-th element of the string
$eig(\rho^{A})$ and
$\lambda^{B}_j$ be the $j$-th element of the string
$eig(\rho^B)$~(one can order the eigenvalues
by an arbitrary way; this ambiguity will not affect the answer).
Let $d^A$ and $d^B$ be the dimensions
of the Hilbert spaces $\calH^A$ and $\calH^B$.
For any bipartite probability distribution
$p^{AB}(i,j)$ of integer variables $i\in [1,d^A]$ and
$j\in[1,d^B]$ denote $p^{AB}$ a string of $d^A\cdot d^B$
probabilities $p^{AB}(1,1),$ $\ldots,
p^{AB}(1,d^B),$ $\ldots p^{AB}(d^A,1),$ $\ldots,p^{AB}(d^A,d^B)$
ordered by an arbitrary way~(this ambiguity will not affect
the answer). In Section~\ref{convex} we will show that
\be
\label{result3}
Conv(M(\lambda)) = \{
(\rho^A,\rho^B) \ : \
\lambda^A_i= \sum_{j=1}^{d^B} p^{AB}(i,j), \ \
\lambda^B_j= \sum_{i=1}^{d^A} p^{AB}(i,j), \ \
p^{AB} \prec \lambda \},
\ee
where $p^{AB}\prec \lambda$ means that the string $p^{AB}$
is majorized by the string $\lambda$~(the majorization
relation is briefly discussed in Section~\ref{convex}).
A set of probability distributions $p^{AB}(i,j)$
satisfying $p^{AB}\prec\lambda$ is known to be a convex one,
and a probability distribution  $p^{AB}(i,j)$ is an
extremal point  if and only if
$p^{AB}=\lambda$ up to some permutation of the elements.
This set of probability distributions is thus a classical
analogue of $Conv(D(\calH,\lambda))$.
The generalization of equality~(\ref{result3}) to the
case of three or more parties is straightforward.

\section{Mean field states for arrays of qubits.}
\label{problem1}

In this section we consider a multipartite
system consisting of qubits (spins $\frac12$) and prove
the equality~(\ref{result1}).
Let $(\rho^{(1)},\ldots,\rho^{(n)})$ be an arbitrary
mean field state of $n$ qubit system. By appropriate unitary local
operators we can always bring the matrices $\rho^{(i)}$ to the
standard form
\be
\label{stand}
\rho^{(i)}=\lambda^{(i)} |1\ra\la 1| + (1-\lambda^{(i)} )
|0\ra\la 0|, \ \
i\in [1,n],
\ee
where $\lambda^{(i)}$  is the {\it lowest} eigenvalue
of $\rho^{(i)}$, that is $\lambda^{(i)}\in [0,\frac12]$.
The proof of~(\ref{result1}) consists of two parts.

\noindent
(a) Suppose that $(\rho^{(1)},\ldots,\rho^{(n)}) \in M_{pure}$, i.e.
$\rho^{(i)} = tr_{j\ne i} (|\Psi\ra\la\Psi|)$ for some pure state
$|\Psi \ra \in (\CC^2)^{\otimes n}$. Let us prove, for example, the
inequality $\lambda^{(n)} \le \sum_{j=1}^{n-1} \lambda^{(j)}$.
If $n$-qubit system is described by the pure state $|\Psi\ra$,
the qubits $1,\ldots,n-1$ alone are described by the density matrix
\be
\rho = tr_n (|\Psi\ra\la \Psi|).
\ee
Taking into account~(\ref{stand}) we can write
\be
\sum_{j=1}^{n-1} \lambda^{(j)}  =
\la \Psi | \sum_{j=1}^{n-1} \Pi_1 [j] |\Psi\ra =
tr(\rho(\sum_{j=1}^{n-1} \Pi_1 [j])),
\ee
where $\Pi_1 = |1\ra\la 1|$ is a one-qubit projector and the
designation $\Pi_1 [j]$ means that the operator $\Pi_1$ is applied
to the $j$-th qubit. Note that the operator $\sum_{j=1}^{n-1} \Pi_1
[j]$ is diagonal in the standard qubit basis and counts a number
of $1$'s in a basis vector. Its only zero eigenvalue corresponds
to the state $|00\ldots 0\ra$. For that reason we will introduce an
auxiliary operator
\be
H = \sum_{j=1}^{n-1} \Pi_1 [j] + |00\ldots 0\ra\la 00\ldots 0|
\ee
Clearly, all eigenvalues of $H$ are greater or equal to unity, i.e.
$H\ge I$. As the whole $n$-qubit system is described by the pure
state $|\Psi\ra$, the positive eigenvalues of the density matrices
$\rho$ and $\rho^{(n)}$ coincide. In particular the highest
eigenvalue of $\rho$ is equal to the highest eigenvalue of
$\rho^{(n)}$ which is $1-\lambda^{(n)}$. It means that $\la
\psi|\rho|\psi\ra\le (1-\lambda^{(n)}) \la \psi|\psi\ra$ for any
$|\psi\ra\in (\CC^2)^{\otimes n-1}$. Thus we conclude that
\be
\sum_{j=1}^{n-1} \lambda^{(j)}  =
tr(\rho H) - \la 00\ldots 0| \rho |00\ldots 0\ra \ge
tr(\rho) - (1 - \lambda^{(n)} ) = \lambda^{(n)},
\ee
which is the desired inequality.

\noindent
(b) We should prove that any mean field state satisfying the
inequalities~(\ref{result1}) does arise from some $n$-qubit
pure state. Obviously one and the same mean field state can arise
from many pure states, so we have some freedom in choosing them.
We will consider only {\it even pure states} (EPS) which are the
states of the following form:
\be
\label{EPS}
|\Psi\ra = \sum_{x\in \BB^n_{even}} c_x |x\ra,
\ \ x=(x_1\ldots x_n), \ \ x_i \in \{ 0,1\}.
\ee
Here $\BB^n_{even}$ is a set  of all binary strings with the
even number of $1$'s. The EPS's span the linear subspace of
dimension $2^{n-1}$.

For any choice of the amplitudes $c_x$ in~(\ref{EPS}) the reduced
density matrices $\rho^{(i)}$ are diagonal in the
$\{ |0\ra, |1\ra \}$ basis and look as
\be
\label{rho^i}
\rho^{(i)} = ( \sum_{x\in \BB^n_{even} \ : \ x_i =0 } |c_x|^2 )
|0\ra\la 0|
+
( \sum_{x\in \BB^n_{even} \ : \ x_i =1} |c_x|^2  ) |1\ra\la 1|,
\ \ i\in [1,n].
\ee
Thus to guarantee that the matrices $\rho^{(i)}$ have the standard
form~(\ref{stand}) we should only require that
\be
\label{highest}
\la 0| \rho^{(i)} |0\ra \ge \la 1| \rho^{(i)} |1\ra,
\ \ i\in [1,n].
\ee
Besides, the set of all mean field states which arise from the EPS's
is a convex one. Indeed, if we take an arbitrary pair of mean field
states $(\rho^{(1)}_0,\ldots,\rho^{(n)}_0)$ and
$(\rho^{(1)}_1,\ldots,\rho^{(n)}_1)$ such that
$$
\rho^{(i)}_{0}  = tr_{j\ne i} ( |\Psi_0 \ra \la \Psi_0 |)
\  { \rm and } \
\rho^{(i)}_{1}  = tr_{j\ne i} (|\Psi_{1}\ra\la \Psi_{1}|),
\ \ i\in [1,n],
$$
for some EPS's
$$
|\Psi_0\ra = \sum_{x\in \BB^n_{even}} c_{x,0} |x\ra \ \ {\rm and} \ \
|\Psi_1\ra = \sum_{x\in \BB^n_{even}} c_{x,1} |x\ra,
$$
then for any $t\in [0,1]$ we can realize
the linear combination $(1-t)(\rho^{(1)}_0,\ldots,\rho^{(n)}_0)
+ t (\rho^{(1)}_1,\ldots,\rho^{(n)}_1)$
by the EPS
$$
|\Psi_t\ra = \sum_{x\in \BB^n_{even}}
\sqrt{ (1-t) |c_{x,0}|^2 + t |c_{x,1}|^2 } \; |x\ra.
$$
Consequently a set of all mean filed states which arise from
EPS's and satisfy the additional requirements~(\ref{highest})
is also a convex one.

The mean field state written in the standard form~(\ref{stand}) can
be identified with a real vector $\vec{\lambda} \in {\RR}^n$
defined as $\vec{\lambda} = (\lambda^{(1)},\ldots,\lambda^{(n)})$.
The previous statement can be rephrased by saying that the vectors
$\vec{\lambda}$ arising from EPS's constitute some convex subset in
$\RR^n$
which we will denote
$\Lambda_{EPS}$.
By definition, $0\le \lambda^{(i)}\le \frac12$, so that
$\Lambda_{EPS}$ is a convex subset of the $n$-dimensional cube $I$
with the edge length $1/2$. Denote $V$ the set of all $2^n$
vertices of $I$, i.e. the vectors $\vec{v} = (\ep_1,\ldots,\ep_n)$,
$\ep \in \{ 0,1/2 \}$. There are exactly $n$ vertices $\vec{v} \in
V$ which violate the inequalities~(\ref{result1}). These are the
vertices containing exactly one $1/2$ component. All other $2^n -
n$ vertices satisfy~(\ref{result1}) and contain either zero or
greater than one $1/2$ components. It will be convenient to
introduce a special designation to distinguish the vertices of the
last type:
\be
\label{V^*}
V^* = \{ \vec{v} \in V \ : \ \vec{v} =
(\ep_1 \ldots \ep_n), \ \
\ep_i \le \sum_{j\ne i }^n \ep_j,
\ \ \ep_i \in \{0,\frac12 \}, \ \
i\in [1,n] \}.
\ee
We claim that any vector from $V^*$ can be realized by an EPS,
i.e. $V^* \subset \Lambda_{EPS}$. Indeed, the vectors
$$
(0\ldots 0), \
(\frac12 \frac12 0 \ldots 0), \
(\frac12 \frac12 \frac12 0 \ldots 0), \ \ldots \
(\frac12 \ldots \frac12) \in V^*
$$
can be realized by the EPS's as follows:
\be
\ba{rcl}
(0\ldots 0) & : & |\Psi\ra = |0\ldots 0\ra,\\ \\
(\frac12 \frac12 0 \ldots 0) & : & |\Psi\ra = 2^{-\frac12}
(|00\ra + |11\ra )\otimes |0\ldots 0\ra, \\ \\
(\frac12 \frac12 \frac12 0 \ldots 0) & : &
|\Psi\ra = 2^{-1} (|000\ra + |011\ra + |101\ra + |110\ra)\otimes
|0\ldots 0\ra, \\
&\cdots & \\
(\frac12 \ldots \frac12) & : &
|\Psi\ra = 2^{-\frac{n-1}2} \sum_{x\in \BB^n_{even}} |x\ra.\\
\ea
\ee
All other vectors from $V^*$ can be realized by EPS's
applying appropriate qubit permutations to the states listed above.
We know that $\Lambda_{EPS}$ is a convex set, so that
any vector from the convex envelope $Conv(V^*)$ also can
be realized by an EPS, i.e.
$Conv(V^*) \subset \Lambda_{EPS}$. To conclude the proof
we will need a simple geometrical lemma which claims that the set
$Conv(V^*)$ coincides with those part of the cube $I$
which is specified by the inequalities~(\ref{result1}):
\be
\label{lemma}
Conv(V^*) = \{ \vec{\lambda} \in I \ : \
\lambda^{(i)} \le \sum_{j\ne i} \lambda^{(j)}, \ \
i\in [1,n] \}.
\ee
We encourage the reader to verify this statement for $n=2,3$
by drawing a picture. The proof of~(\ref{lemma}) for an
arbitrary $n$ is given in the Appendix~A. Thus for any
$\vec{\lambda} \in I$ such that $\lambda^{(i)} \le
\sum_{j\ne i} \lambda^{(j)}$, $i \in [1,n]$ we have proven
that $\vec{\lambda} \in Conv(V^*) \subset
\Lambda_{EPS}$, so that $\vec{\lambda}$
can be realized by the EPS. It means that any mean field state
satisfying~(\ref{result1}) can be realized by some
$n$-qubit pure state.

Note that~(\ref{lemma}) implies also a converse inclusion
$\Lambda_{EPS} \subset Conv(V^*)$ because we already know that
inequalities~(\ref{result1}) are necessary for compatibility
of the mean field state with a pure state.
It means that $\Lambda_{EPS} = Conv(V^*)$.

\section{Mean field states for two qubits}
\label{problem2}

In this section we consider a bipartite system composed of two qubits
and prove that the inequalities~(\ref{result2}) are necessary
for compatibility of a mean field state $(\rho^A,\rho^B)$
with at least one  mixed state from $D(\CC^4,\lambda)$.
A proof that~(\ref{result2}) is also a sufficient condition
for compatibility is placed in Appendix~B.

The combinations of $\lambda_i$'s standing in the righthand sides
of the inequalities~(\ref{result2})
can be easily found if we consider two {\it
classical} bits and joint probability distributions of two bits
with the specified set of probabilities
$\{\lambda_1,\lambda_2,\lambda_3,\lambda_4\}$ which can be
arbitrarily assigned to the events 00, 01, 10, and 11
(this is a classical
analog of the set $D(\CC^4,\lambda)$). Generally there are $4!$
such distributions. In this case the reduced density matrices
$\rho^A$, $\rho^B$ become the partial one-bit probability distributions
and $\lambda^A$, $\lambda^B$ become the minimal values of $\rho^A$,
$\rho^B$. In these settings the inequalities~(\ref{result2}) are
the best estimates on the quantities $\lambda^A$, $\lambda^B$,
$\lambda^A + \lambda^B$, and $|\lambda^A - \lambda^B|$ as can be
explicitly verified. The nontrivial result is that this estimates
also hold in the quantum case and the set $M(\lambda)$ is
completely specified by them.

As we will see below, the strict proof of the
inequality $|\lambda^A- \lambda^B| \le \min \{ \lambda_1 -
\lambda_3, \lambda_2 - \lambda_4 \}$  is rather
formidable. However in one particular case it can be proved without
any calculations. Suppose that the minimum here is zero, i.e.
$\lambda_1=\lambda_3$ or $\lambda_2=\lambda_4$. Then the spectrum
$\lambda$ involves an eigenvalue which is degenerated with
multiplicity three or four. In this case we can represent any
density matrix $\rho\in D(\CC^4,\lambda)$ as a linear combination
of the identity operator and a projector on some pure state.
Obviously the term proportional to identity operator just shifts
the spectrums of $\rho^A$ and $\rho^B$ by the same constant. So it
doesn't contribute to the combination $\lambda^A - \lambda^B$. The
term proportional to a projector on a pure state produces the
reduced density matrices $\rho^A$ and $\rho^B$ with the same
spectrum according to the Schmidt theorem. So we conclude that
$\lambda^A=\lambda^B$.

Suppose that $(\rho^A,\rho^B) \in M(\lambda)$, i.e.
$\rho^A = tr_B(\rho)$, $\rho^B = tr_A(\rho)$ for some
$\rho\in D(\CC^4,\lambda)$. We should prove that
$\lambda^A$, $\lambda^B$, and $\lambda$ satisfy
four inequalities~(\ref{result2}).
 Let us represent the matrices
$\rho^A$, $\rho^B$ as
\be
\rho^A = \lambda^A \Pi^A + (1-\lambda^A) (I- \Pi^A), \ \
\rho^B = \lambda^B \Pi^B + (1-\lambda^B) (I- \Pi^B),
\ee
where $\Pi^A$, $\Pi^B$ are the projectors on the eigenstates
of $\rho^A$, $\rho^B$ with the lowest eigenvalue, so that
$\lambda^A= tr(\rho \Pi^A[A])$, $\lambda^B= tr(\rho \Pi^B[B])$.
As usual the qubit acted on by  one-qubit operator is indicated
in the square brackets.

Suppose now that $O$ is an arbitrary two-qubit hermitian operator
with the eigenvalues $O_1\ge O_2\ge O_3\ge O_4$. We claim that
\be
\label{infO}
\inf_{\eta \in D(\CC^4,\lambda)}
tr(O\eta) =
\lambda_1 O_4 +\lambda_2 O_3 + \lambda_3 O_2 + \lambda_4 O_1.
\ee
To verify this statement one can consider the variation $\delta
\eta = i[h,\eta]$ where $h^\dag = h$ is an arbitrary infinitesimal
hermitian operator. This variation is obviously consistent with the
constraint $\eta \in D(\CC^4,\lambda)$. Suppose that the minimum
in~(\ref{infO}) is achieved at the matrix $\eta^* \in
D(\CC^4,\lambda)$. Then we have
\be
\delta tr(O\eta^*) = i tr(O[h,\eta^*]) = i tr(h [\eta^*,O]) =0
\ee
for any $h$. It means that $[\eta^*,O]=0$ so that $\eta^*$ and $O$
are diagonal in the same basis. Therefore there is a permutation
$\sigma \in S_4$ such that
\be
tr(O\eta^*) =
\lambda_1 O_{\sigma(4)} +\lambda_2 O_{\sigma(3)}
 + \lambda_3 O_{\sigma(2)} + \lambda_4 O_{\sigma(1)}.
\ee
It is easy to verify that the optimal permutation is always
$\sigma(i) = i$
and thus the equality~(\ref{infO}) is proven.

Let us substitute $O=\Pi^A[A]$ into the equality~(\ref{infO}).
In this case we have $\{ O_1,O_2,O_3,O_4\}=\{1,1,0,0\}$ and thus
\be
\lambda^A = tr(\rho \Pi^A[A]) \ge
\inf_{\eta \in D(\CC^4,\lambda)}
tr(\eta \Pi^A[A]) = \lambda_4 + \lambda_3.
\ee
By the same way substituting $O=\Pi^B[B]$ and
$O=\Pi^A[A] + \Pi^B[B]$ into~(\ref{infO}) we get the inequalities
$\lambda^B\ge \lambda_4 + \lambda_3$ and
$\lambda^A + \lambda^B \ge 2\lambda_4 + \lambda_2 + \lambda_3$.

Unfortunately the last inequality of~(\ref{result2}), that is
\be
\label{estimate}
|\lambda^A - \lambda^B| \le
\min \{ \lambda_1 - \lambda_3, \lambda_2 - \lambda_4 \},
\ee
can not be proved using the same idea (the analogous method
allows to get the estimate $|\lambda^A - \lambda^B| \le
\lambda_1 - \lambda_4$ which is too weak).
The proof of~(\ref{estimate}) presented below is a bit formidable.

Let us define polarization vectors for the qubits A and B:
\be
\label{polar}
a^\alpha[\eta] = tr(\sigma^\alpha[A] \eta), \ \
b^\alpha[\eta] = tr(\sigma^\alpha[B] \eta),
\ee
where $\sigma^x$, $\sigma^y$, $\sigma^z$ are the Pauli matrices
and $\eta\in D(\CC^4)$ is an arbitrary density matrix. Denote also
$a=\sqrt{a^\alpha a^\alpha}$ and $b=\sqrt{b^\alpha b^\alpha}$ the
absolute magnitudes of the polarization vectors. We
can express the eigenvalues $\lambda^A$ and $\lambda^B$ in terms
of polarizations as $\lambda^A=\frac12 (1-a[\rho])$ and
$\lambda^B = \frac12 (1-b[\rho])$. Introduce a functional
\be
\label{F}
F[\eta] = b[\eta] - a[\eta], \ \
\eta \in D(\CC^4),
\ee
such that $|\lambda^A- \lambda^B| = \frac12 |F[\rho]|$. The
inequality~(\ref{estimate}) is equivalent to
the following one:
\be
\label{supF}
\sup_{\eta \in D(\CC^4,\lambda)} F[\eta] \le 2
\min \{ \lambda_1 - \lambda_3, \lambda_2 - \lambda_4 \}
\ee
Note that an application of the qubit swap transformation changes the sign of
$F$ and doesn't changes the eigenvalues of $\eta$ so the
supremum~(\ref{supF}) can not be negative. Let $\eta^* \in
D(\CC^4,\lambda)$ be some density matrix at which this supremum is
achieved (may be it is not unique). Consider the variation $\delta
\eta = i[h,\eta^*]$ where $h^\dag = h$ is an arbitrary
infinitesimal hermitian operator. This variation is consistent with
the constraint $\eta \in D(\CC^4,\lambda)$. We must require that
$\delta F[\eta^*]=0$.

Suppose first that both polarization vectors $a^\alpha \equiv
a^\alpha[\eta^*]$ and $b^\alpha \equiv b^\alpha[\eta^*]$ are
non-zero. Denote $\hat{a}^\alpha=a^\alpha/a$ and
$\hat{b}^\alpha=b^\alpha/b$ the normalized polarization vectors.
The equation $\delta F[\eta^*]=0$ then can be written as
\be
tr(h\left[ \eta^*,\hat{a}^\alpha \sigma^\alpha[A] -
\hat{b}^\alpha \sigma^\alpha[B] \right])= 0.
\ee
As $h$ may be an arbitrary hermitian operator we conclude that
\be
\label{extr}
\hat{a}^\alpha [\sigma^\alpha[A],\eta^*] =
\hat{b}^\alpha [\sigma^\alpha[B],\eta^*].
\ee
Let us parameterize the density matrix $\eta^*$ as follows:
\be
\label{parametr}
\eta^* = \frac14 (
I + a^\alpha \sigma^\alpha[A] + b^\alpha \sigma^\alpha[B] +
A_{\alpha,\beta}\, \sigma^\alpha[A] \sigma^\beta[B]),
\ee
where $A_{\alpha,\beta}$ is some 3x3 real matrix. Using the
commutators $[\sigma^\alpha,\sigma^\beta]=
2i\ep^{\alpha\beta\gamma}\sigma^\gamma$
we can transform the equation~(\ref{extr}) into
\be
\label{extr1}
\hat{a}^\gamma A_{\delta,\alpha} \ep^{\gamma\delta\beta} =
\hat{b}^\gamma A_{\beta,\delta} \ep^{\gamma\delta\alpha}.
\ee
As the functional $F[\eta]$ is invariant under one-qubit unitary
transformations, we can assume that both qubits are polarized
along $z$-axis, i.e. $\hat{a}=\hat{b}=(0,0,1)$. Under this
assumption one can easily check that the most common solution
of the equation~(\ref{extr1}) looks as
\be
A=\left[ \ba{ccc} s & t  & 0 \\
                  t & -s & 0 \\
                  0 & 0  & r \\
         \ea \right]
\ee
for some real numbers $s,t,r$. Substituting it into~(\ref{parametr})
we can represent the matrix $\eta^*$ in the basis
$\{ |00\ra,|01\ra,|10\ra,|11\ra\}$ of $\CC^4$ as follows
\be
\eta^* = \frac14 \left[ \ba{cccc}
1+ a + b + r & 0 & 0 & \bar{z} \\
0 & 1 + a - b - r & 0 & 0 \\
0 & 0 & 1 - a + b - r & 0 \\
z & 0 & 0 & 1 - a - b + r \\
\ea \right],
\ee
where $z=2(s+it)$. Note that $\eta^*$ is a block diagonal matrix
so its eigenvalues
$\{\lambda^*_1,\lambda^*_2,\lambda^*_3,\lambda^*_4\}$ can be easily
found:
\be
\label{l*}
\ba{rcl}
\lambda^*_1 &=& \frac14 ( 1 + a - b - r), \\ \\
\lambda^*_2 &=& \frac14 ( 1 - a + b - r), \\
\ea \ \
\ba{rcl}
\lambda^*_3 &=& \frac14 ( 1 + r + \sqrt{(a+b)^2 + |z|^2} ), \\ \\
\lambda^*_4 &=& \frac14 ( 1 + r - \sqrt{(a+b)^2 + |z|^2} ). \\
\ea
\ee
Recall that $\eta^* \in D(\CC^4,\lambda)$ so that the eigenvalues
$\lambda^*_i$ coincide up to some permutation with the eigenvalues
$\lambda_i$. Although this permutation somehow depends upon
$a,b,r,z$ (recall that the eigenvalues $\lambda_i$ are ordered in
decreasing way) we need not to consider all possible cases. Instead
we can notice that
\be
\max_{\sigma \in S^4} \;
\min\{ |\lambda_{\sigma(1)} - \lambda_{\sigma(3)}|,
       |\lambda_{\sigma(2)} - \lambda_{\sigma(4)}|\} =
\min\{ \lambda_1 - \lambda_3, \lambda_2 - \lambda_4 \},
\ee
where the maximum is taken over all permutations of four elements.
Note also that the lefthand side is invariant under the substitution
$\lambda_i \to \lambda^*_i$. Thus we conclude that
\be
\min\{ |\lambda^*_1 - \lambda^*_2|, |\lambda^*_3 - \lambda^*_4|\} \le
\min\{ \lambda_1 - \lambda_3, \lambda_2 - \lambda_4 \}.
\ee
Taking $\lambda^*_i$ from the table~(\ref{l*}) we arrive to
the inequality
\be
\frac12 \min\{ | a - b|, \sqrt{ (a+b)^2 + |z|^2} \} \le
\min\{ \lambda_1 - \lambda_3, \lambda_2 - \lambda_4 \}.
\ee
The minimum in the lefthand side is equal to $|a-b|$
regardless of the parameter $z$. Consequently
\be
b-a \le 2 \min\{ \lambda_1 - \lambda_3, \lambda_2 - \lambda_4 \},
\ee
which proves the inequality~(\ref{supF}).

Suppose now that at the supremum of $F[\eta]$
one of the polarization vectors becomes zero.
The only non-trivial situation is
$a^\alpha[\eta^*]=0$. In this case the requirement $\delta
F[\eta^*]=0$ under the variation $\delta \eta =i[h,\eta^*]$ implies
that $\delta a^\alpha[\eta^*] =0$ for all three components
$\alpha$. By analogy with~(\ref{extr}) this requirement implies
that
\be
[\sigma^\alpha[A],\eta^*]=0, \ \ \alpha=x,y,z.
\ee
It means that $\eta^*$ commutes with any operator acting
only on the
qubit A. This is possible only if $\eta^*$ has the form
$\eta^* = (I/2)\otimes \eta^B$ for some one-qubit density matrix
$\eta^B$. Thus the eigenvalues of $\eta^*$ are
\be
\lambda_1 = \lambda_2 = \frac14 (1 + b), \ \
\lambda_3 = \lambda_4 = \frac14 (1 - b).
\ee
Note that $\eta^*\in D(\CC^4,\lambda)$ only if $\lambda_1 =
\lambda_2$ and $\lambda_3= \lambda_4$ so the considered situation
is not a generic one. Nevertheless we have $F[\eta^*] = b =
2(\lambda_1 - \lambda_3) = 2(\lambda_2 - \lambda_4)$ so that the
estimate~(\ref{supF}) is valid in this case also.

\section{Connection with multipartite probability distributions}
\label{convex}

In this section we consider a bipartite system with  local
Hilbert spaces $\calH^A$ and $\calH^B$ of arbitrary dimensions.
We show that a set
\be
\label{conv1}
Conv(M(\lambda)) = \{ (\rho^A,\rho^B) \ : \
\rho^A= tr_B(\rho), \ \rho^B = tr_A(\rho), \
\rho\in Conv(D(\calH,\lambda)) \}
\ee
can be described in terms of multipartite
probability distributions according to~(\ref{result3}).
The restriction that the system is bipartite is not essential
and is needed only to simplify the designations.
Using the arguments presented below, one can easily
formulate and prove a multipartite analogue of~(\ref{result3}).
We start by collecting all relevant facts concerning the majorization
relation.

Consider  arbitrary strings of non-negative real numbers
$p=(p_1,\ldots,p_d)$ and $q=(q_1,\ldots,q_d)$ such that
$\sum_{i=1}^d p_i = \sum_{i=1}^d q_i  =1$.
Let us arrange the elements of $p$ and $q$ in decreasing order.
Denote $p^\da_i$ the $i$-th element of $p$ from above and
$q^\da_i$ the $i$-th element of $q$ from above, such that
$p^\da_1\ge p^\da_2\ge\cdots\ge p^\da_d$ and
$q^\da_1\ge q^\da_2\ge\cdots\ge q^\da_d$.
The string $p$ is said to be majorized by the string $q$,
if for each $k$ in the range $1,\ldots,d-1$ we have
\be
\label{maj}
\sum_{i=1}^k p^\da_i \le \sum_{i=1}^k q^\da_i.
\ee
This relation is denoted as $p\prec q$. We will need
three well known facts given below:

\noindent
(i)~$p\prec q$ if and only if
$p_i = \sum_{\sigma \in S_d} t_\sigma q_{\sigma(i)}$,
$i=1,\ldots,d$, where $S_d$ is the symmetric group,
$\sigma \in S_d$ is a permutation of $d$ elements,
$t_\sigma\ge0$ and $\sum_{\sigma\in S_d} t_\sigma=1$.
The existence of such decomposition is guaranteed by
Birkhoff and von Neumann decomposition of
a doubly stochastic matrix into a convex
combination of permutation matrices.

\noindent
(ii) Let $\calS$ be an arbitrary Hilbert space.
Consider any pair of density matrices $\rho,\eta \in
D(\calS)$. Then $eig(\rho)\prec eig(\eta)$ if and only if $\rho=
\sum_\alpha t_\alpha U_\alpha \eta U_\alpha^\dag$, where $U_\alpha$
are some unitary operators on $\calS$, $t_\alpha\ge 0$ and
$\sum_\alpha t_\alpha=1$, see~\cite{Werhl}.

\noindent
(iii) Let $\calS$ be an arbitrary Hilbert space.
For any $\rho\in D(\calS)$ and for any orthonormal basis
in $\calS$, the string of diagonal elements of $\rho$
is majorized by the eigenvalues of $\rho$, i.e.
$diag(\rho)\prec eig(\rho)$.

Note that according to the property~(ii), the requirement
$\rho\in Conv(D(\calH,\lambda))$ is equivalent to
$\rho\in D(\calH)$ and $eig(\rho)\prec \lambda$,
so we can rewrite~(\ref{conv1}) as
\be
\label{conv2}
Conv(M(\lambda)) = \{ (\rho^A,\rho^B) \ : \
\rho^A= tr_B(\rho), \ \rho^B = tr_A(\rho), \
\rho \in D(\calH),  \ eig(\rho)\prec \lambda \}.
\ee
Now we are ready to prove the result~(\ref{result3}).
The proof consists of two parts.\\
(a) Let  $(\rho^A,\rho^B)$ be an arbitrary mean field state
such that
\be
\label{average}
\lambda^A_i =\sum_{j=1}^{d^B} p^{AB}(i,j), \ \
\lambda^B_j=\sum_{i=1}^{d^A} p^{AB}(i,j),
\ee
for some bipartite probability distribution $p^{AB}(i,j)$
satisfying $p^{AB} \prec \lambda$.
Consider the eigenvectors decomposition
of the density matrices $\rho^A$ and $\rho^B$:
\be
\label{Abdecomp}
\rho^A=\sum_{i=1}^{d^A} \lambda^A_i |i_A\ra\la i_A|, \ \
\rho^B=\sum_{j=1}^{d^B} \lambda^B_j |j_B\ra\la j_B|,
\ee
where $|i_A\ra \in \calH^A$, $|j_B\ra \in \calH^B$.
Define a bipartite density matrix $\rho \in D(\calH)$
according to
\be
\rho=\sum_{i=1}^{d^A}\sum_{j=1}^{d^B}
p^{AB}(i,j)|i_A\ra \la i_A| \otimes
|j_B\ra\la j_B|.
\ee
By definition, $\rho^A = tr_B(\rho)$, $\rho^B =tr_A(\rho)$,
and $eig(\rho)=p^{AB}\prec\lambda$, so that
$(\rho^A,\rho^B)\in Conv(M(\lambda))$, see~(\ref{conv2}).\\
(b) Now suppose that $(\rho^A,\rho^B)\in Conv(M(\lambda))$, i.e.
that $\rho^A = tr_B(\rho)$, $\rho^B = tr_A(\rho)$ for some $\rho
\in D(\calH)$ such that $eig(\rho)\prec \lambda$,
see~(\ref{conv2}). Consider the eigenvectors
decomposition~(\ref{Abdecomp}) and define a bipartite probability
distribution $p^{AB}(i,j)$ according to
\be
p^{AB}(i,j)=\la i_A,j_B|\rho | i_A,j_B\ra,\ \
i\in [1,d^A],\ j\in [1,d^B].
\ee
Clearly, this probability
distribution satisfies the
requirements~(\ref{average}). By definition, in the basis
$\{|i_A,j_B\ra\}_{i,j}$ we have
$p^{AB}=diag(\rho)$, so the property~(iii) tells us that
$p^{AB}\prec eig(\rho)$. Summarizing, we have: $p^{AB}\prec
eig(\rho)\prec\lambda$. Transitivity of majorization relation
implies that $p^{AB}\prec \lambda$. This completes the proof
of the equality~(\ref{result3}).

For an arbitrary string of eigenvalues $\lambda$, let us
denote $P(\lambda)$ the set of all bipartite probability
distributions  $p^{AB}(i,j)$  such that $p^{AB}=\lambda$
up to some permutation. The set $P(\lambda)$ is a classical
analogue of the set $D(\calH,\lambda)$.
The property~(i) tells us that
$p^{AB} \prec \lambda$ if and only if
$p^{AB} \in Conv(P(\lambda))$. Thus the reduction of the
Problem~3 to the analogous classical problem
corresponds to the substitutions $(\rho^A,\rho^B)
\to (\lambda^A_i,\lambda^B_j)$, $\rho \to p^{AB}(i,j)$,
$D(\calH,\lambda) \to P(\lambda)$.
The partial traces corresponds to taking the average over
one variable.

The inclusion $(\rho^{(1)},\ldots,\rho^{(n)})\in Conv(M(\lambda))$
may be regarded as a necessary condition for
inclusion $(\rho^{(1)},\ldots,\rho^{(n)})\in M(\lambda)$.
It provides us some knowledge about the sets $M(\lambda)$
and $M_{pure}$ for an arbitrary multipartite system.
For example consider a system of three particles A,B, and C.
Let $(\rho^A,\rho^B,\rho^C)$ be an arbitrary mean field state.
The requirements
\be
\label{ABC}
\ba{rcl}
(\rho^A,\rho^B) &\in& Conv(M(eig(\rho^C))), \\
(\rho^B,\rho^C) &\in& Conv(M(eig(\rho^A))), \\
(\rho^C,\rho^A) &\in& Conv(M(eig(\rho^B)))
\ea
\ee
are necessary for inclusion $(\rho^A,\rho^B,\rho^C)
\in M_{pure}$. If A,B, and C are the qubits, one
can check that these requirements are also sufficient.
Meanwhile if $\calH^A=\calH^B=\CC^2$ and
$\calH^C=\CC^4$, one can build an explicit example
which shows that they are not sufficient.
Indeed, take $eig(\rho^A)=\{0.6, 0.4\}$,
$eig(\rho^B)=\{0.5, 0.5\}$, and
$eig(\rho^C)=\{0.3, 0.3, 0.3, 0.1\}$.
One can verify that $eig(\rho^A\otimes\rho^B)\prec eig(\rho^C)$,
$eig(\rho^B\otimes\rho^C)\prec eig(\rho^A)$, and
$eig(\rho^C\otimes\rho^A)\prec eig(\rho^B)$, so that~(\ref{ABC})
are satisfied. However $(\rho^A,\rho^B,\rho^C)\notin M_{pure}$
because there is another necessary requirement which is
violated by this mean field state, see Section~\ref{problem2}.

\section{Conclusions}
We found the necessary and sufficient conditions under which
a collection of density matrices~(a mean field state)
can be produced by the partial trace operations from a multipartite
pure state for the system of $n$ qubits and for the system
$\CC^2\otimes\CC^2\otimes\CC^4$. These conditions may be described
by linear inequalities on the eigenvalues of local 
density matrices. We also found the necessary and sufficient
conditions under which a mean field state can be produced by the
partial trace operations from a multipartite mixed state whose
eigenvalues are majorized by a given real vector.

\section{Acknowledgements}
I would like to thank David DiVincenzo and Alexei Kitaev
for advices on majorization technique. I am also very grateful to 
Alexei Kitaev for thoroughly  reading the manuscript
and for many suggestions which helped to improve the
paper.
This work was supported by NWO-Russia
collaboration program.

\section{Appendix A}
The purpose of this section is to prove the equality~(\ref{lemma}).
Denote $\hat{I}$ the subset of the cube $I$ which is specified
by the inequalities~(\ref{result1}), i.e.
\be
\label{conv}
\hat{I} = \{ \vec{\lambda} \in \RR^n \ : \
\lambda^{(i)} \le \sum_{j\ne i} \lambda^{(j)}, \ \
0\le \lambda^{(i)} \le \frac12, \ \
i\in [1,n]\}.
\ee
Clearly $I$ is a convex set.
The extremal
point of $\hat{I}$ is a vector $\vec{\lambda} \in \hat{I}$ for
which some $m\ge n$ of the inequalities~(\ref{conv}) specifying
$\hat{I}$ become the equalities and the corresponding system of $m$
equations has a unique solution (thus having a rank $n$).
Let us prove that if $\vec{\lambda}\in \hat{I}$ is an extremal
point of $\hat{I}$, then $\vec{\lambda}\in V^*$. Because any
convex set is a convex envelope of its extremal points, it will
imply that $\hat{I} \subset Conv(V^*)$.

To each extremal point $\vec{\lambda} \in \hat{I}$ we can assign a
pair of integers $(p,q)$ where $p\in [0,n]$ is the number of
equalities $\lambda^{(i)} = \sum_{j\ne i} \lambda^{(j)}$ which
are satisfied at $\vec{\lambda}$ and $q\in [0,n]$ is the number
of equalities $\lambda^{(i)}=0$, $\lambda^{(i)}=\frac12$
which are satisfied at $\vec{\lambda}$. By definition $p+q\ge n$
but we can assume that $p+q=n$ because at each extremal point
there are only $n$ independent equations.
Now we are ready to list all extremal points of $\hat{I}$ and
check that they all belong to $V^*$.

\noindent
(a) The extremal points of the type $(p,n-p)$, $p\ge 3$ \\ We have
the equalities $\lambda^{(i)} = \sum_{j\ne i} \lambda^{(j)}$,
$\lambda^{(k)} = \sum_{j\ne k} \lambda^{(j)}$, $\lambda^{(l)} =
\sum_{j\ne l} \lambda^{(j)}$ for some $i\ne k \ne l$. They are
compatible with the inequalities $\lambda_j \ge 0$, $j\in [1,n]$
only if $\vec{\lambda} = (0\ldots 0)$.

\noindent
(b) The extremal points of the type $(2,n-2)$. \\
The equalities $\lambda^{(i)} = \sum_{j\ne i} \lambda^{(j)}$,
$\lambda^{(k)} = \sum_{j\ne k} \lambda^{(j)}$ and
the inequalities $\lambda_j \ge 0$, $j\in [1,n]$ are
compatible only if $\lambda^{(i)}=\lambda^{(k)}$
and $\lambda^{(j)}=0$ for $j\ne i$ or $j\ne k$. Thus we
have either $\lambda^{(i)}=\lambda^{(k)}=0$ or
$\lambda^{(i)}=\lambda^{(k)}=\frac12$. The extremal points
of both types belongs to $V^*$.

\noindent
(c) The extremal points of the type $(1,n-1)$. \\ Suppose that
$\lambda^{(i)}=\sum_{j\ne i} \lambda^{(j)}$ for some $i\in [1,n]$.
If we choose the equality $\lambda^{(i)}=0$ then $\lambda^{(j)}=0$
for all $j\ne i$. If we choose the equality $\lambda^{(i)}=\frac12$
then we must choose $\lambda^{(k)}=\frac12$ for some $k \ne i$ and
$\lambda^{(j)}=0$ for $j\ne i$ and $j\ne k$. If we do not choose
neither $\lambda^{(i)}=0$ nor $\lambda^{(i)}=\frac12$ then for each
$j\ne i$ we must choose either $\lambda^{(j)}=0$ or
$\lambda^{(j)}=\frac12$. This choice is compatible with the
equality $\lambda^{(i)}=\sum_{j\ne i} \lambda^{(j)}$ only if
$\lambda^{(i)}=0$ or $\lambda^{(i)}=\frac12$. In all these cases we
arrive to the extremal points which have been already listed in (a)
or (b).

\noindent
(d)  The extremal points of the type $(0,n)$. \\
The extremal point of this type is a vertex of the cube $I$
which satisfies the inequalities~(\ref{result1}). According
to definition~(\ref{V^*}) any point of this type belongs
to $V^*$.

So  all extremal points of $\hat{I}$ belong to $V^*$ and thus
$\hat{I} \subset Conv(V^*)$. The inclusion $Conv(V^*)
\subset \hat{I}$ is obvious
because $V^* \subset \hat{I}$ and $\hat{I}$ is a convex set.
Thus the equality $\hat{I}=Conv(V^*)$ is proven.

\section{Appendix B}

In this section we show that if a mean field state of two qubits
$(\rho^A,\rho^B)$ satisfies the inequalities~(\ref{result2}) for
some eigenvalue string
$\lambda=(\lambda_1,\lambda_2,\lambda_3,\lambda_4)$, then it can be
represented as $\rho^A =tr_B(\rho)$, $\rho^B=tr_A(\rho)$ for some
two qubit density matrix $\rho\in D(\CC^4,\lambda)$, i.e. that
$(\rho^A,\rho^B)\in M(\lambda)$. It is convenient to draw a
region $\Omega$ defined by the inequalities~(\ref{result2})
on the plane $(\lambda^A,\lambda^B)$ for a fixed string $\lambda$,
see FIG.~1.
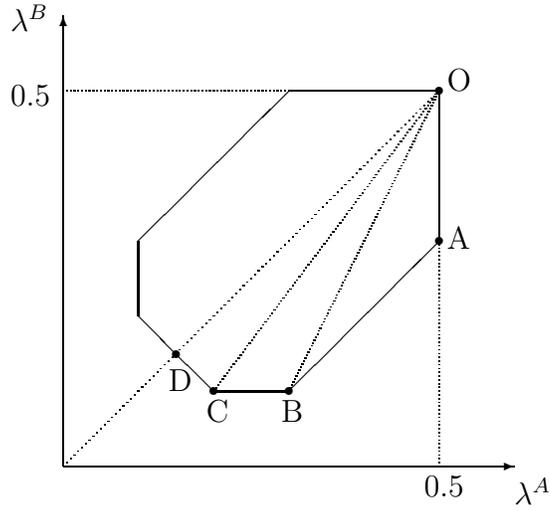
\begin{figure}
\unitlength=1mm
\begin{center}
\begin{picture}(60,60)
\put(0,0){\line(1,1){2}}
\put(0,0){\vector(1,0){60}}
\put(0,0){\vector(0,1){60}}
\multiput(50,0)(0,0.5){100}{\line(0,1){0.05}}
\multiput(0,50)(0.5,0){100}{\line(0,1){0.05}}
\multiput(0,0)(0.5,0.5){100}{\line(0,1){0.05}}
\put(20,10){\line(-1,1){10}}
\put(20,10){\line(1,0){10}}
\put(30,10){\line(1,1){20}}
\put(50,30){\line(0,1){20}}
\put(10,20){\line(0,1){10}}
\put(10,30){\line(1,1){20}}
\put(30,50){\line(1,0){20}}
\put(60,-5){$\lambda^A$}
\put(-7,58){$\lambda^B$}
\put(48,-4){$0.5$}
\put(-7,48){$0.5$}
\put(20,10){\circle*{1}}
\put(19,6){C}
\put(30,10){\circle*{1}}
\put(29,6){B}
\put(50,30){\circle*{1}}
\put(51,29){A}
\put(50,50){\circle*{1}}
\put(51,50){O}
\put(15,15){\circle*{1}}
\put(14,10){D}
\multiput(20,10)(0.3,0.4){100}{\line(0,1){0.05}}
\multiput(30,10)(0.2,0.4){100}{\line(0,1){0.05}}
\end{picture}
\end{center}
\caption{
The region $\Omega$ on the plane $(\lambda^A,\lambda^B)$ specified
by the inequalities~(\ref{result2}) for $\lambda=(0.7,0.2,0.1,0)$.
For an arbitrary $\lambda$ the coordinates of the vertices $O$,
$A$, $B$, $C$, $D$ are: $O=(\frac12,\frac12)$, $A=(\frac12,\frac12
- \min{[\lambda_1 - \lambda_3,
\lambda_2 - \lambda_4]})$,
$B=(\min{[\lambda_1 + \lambda_4,\lambda_2 + \lambda_3]},
\lambda_3 + \lambda_4)$,
$C=(\lambda_2+\lambda_4,\lambda_3 + \lambda_4)$,
$D=\frac12 (2\lambda_4 + \lambda_3 +
\lambda_2, 2\lambda_4 + \lambda_3 + \lambda_2)$.}
\end{figure}
We will describe the desired matrix $\rho\in D(\CC^4,\lambda)$
explicitly for each point inside $\Omega$. It is sufficient to
consider only the case $\lambda^A\ge \lambda^B$ because the
constraint $\rho\in D(\CC^4,\lambda)$ is invariant under the
permutation of the qubits. Also we will assume that $\lambda_1 -
\lambda_3\ge \lambda_2 - \lambda_4$. The opposite case may be
investigated similarly. We will describe three independent families
of the density matrices $\rho\in D(\CC^4,\lambda)$ corresponding to
three triangles OCD, OAB, and OBC on FIG.~1. Let us define two
qubit states
\be
\ba{rcl}
|\Psi_1\ra &=& 2^{-\frac12}
\left[\sqrt{1-\alpha}\; |0,0\ra
+ \sqrt{1+\alpha}\; |1,1\ra\right],\\ \\
|\Psi_2\ra &=& 2^{-\frac12}
\left[\sqrt{1+\alpha}\; |0,0\ra
- \sqrt{1-\alpha}\; |1,1\ra\right],\\ \\
|\Psi_3\ra &=& 2^{-\frac12}
\left[\sqrt{1-\beta}\; |0,1\ra
+ \sqrt{1+\beta}\; |1,0\ra\right],\\ \\
|\Psi_4\ra &=& 2^{-\frac12}
\left[\sqrt{1+\beta}\; |0,1\ra
- \sqrt{1-\beta}\; |1,0\ra\right].\\
\ea
\ee
For any parameters $\alpha,\beta\in [-1,1]$ this is an orthonormal
basis in $\CC^4$, i.e. $\la\Psi_i|\Psi_j\ra=\delta_{i,j}$. Consider
the density matrix
\be
\rho=\lambda_1 |\Psi_1\ra\la\Psi_1| +
\lambda_2 |\Psi_4\ra\la\Psi_4| +
\lambda_3 |\Psi_3\ra\la\Psi_3| +
\lambda_4 |\Psi_2\ra\la\Psi_2| \in D(\CC^4,\lambda).
\ee
The reduced density matrices $\rho^A=tr_B(\rho)$ and
$\rho^B=tr_A(\rho)$ have the lowest eigenvalues
\be
\lambda^A=\frac12
\left\{ 1- \left| \alpha(\lambda_1 - \lambda_4) -
\beta (\lambda_2 - \lambda_3)\right| \right\},
\ \
\lambda^B=\frac12
\left\{ 1- \left| \alpha(\lambda_1 - \lambda_4) +
\beta (\lambda_2 - \lambda_3)\right| \right\}.
\ee
One can easily verify that the region of parameters $0\le \alpha
\le 1$, $0\le \beta \le \alpha$ is mapped to the triangle OCD on
FIG.~1. Next consider the density matrix
\be
\rho=\lambda_1 |\Psi_1\ra\la\Psi_1| +
\lambda_2 |\Psi_4\ra\la\Psi_4| +
\lambda_3 |\Psi_2\ra\la\Psi_2| +
\lambda_4 |\Psi_3\ra\la\Psi_3| \in D(\CC^4,\lambda).
\ee
The reduced density matrices $\rho^A=tr_B(\rho)$ and
$\rho^B=tr_A(\rho)$ have the lowest eigenvalues
\be
\lambda^A=\frac12
\left\{ 1- \left| \alpha(\lambda_1 - \lambda_3) -
\beta (\lambda_2 - \lambda_4)\right| \right\},
\ \
\lambda^B=\frac12
\left\{ 1- \left| \alpha(\lambda_1 - \lambda_3) +
\beta (\lambda_2 - \lambda_4)\right| \right\}.
\ee
One can easily verify that the region of parameters $0\le \beta \le
1$, $\frac{\lambda_2-\lambda_4}{\lambda_1 - \lambda_3}\beta \le
\alpha \le \beta$ is mapped to the triangle OAB on FIG.~1.
Now let us define two qubit states
\be
\ba{rcl}
|\tilde{\Psi}_1\ra &=& 2^{-\frac12}
\left[\sqrt{1-\alpha}\; |0,0\ra +
\sqrt{1+\alpha}\; |1,1\ra\right],\\ \\
|\tilde{\Psi}_2\ra &=& 2^{-\frac12}
\left[\sqrt{1+\alpha}\; |\phi_0,0\ra -
\sqrt{1-\alpha}\; |M_1,1\ra\right],\\ \\
|\tilde{\Psi}_3\ra &=& 2^{-\frac12}
\left[\sqrt{1-\alpha}\; |\phi_0,1\ra +
\sqrt{1+\alpha}\; |\phi_1,0\ra\right],\\ \\
|\tilde{\Psi}_4\ra &=& 2^{-\frac12}
\left[\sqrt{1+\alpha}\; |0,1\ra -
\sqrt{1-\alpha}\; |1,0\ra\right],\\
\ea
\ee
where $|\phi_0\ra = \cos{(\frac{\varphi}2)}|0\ra +
\sin{(\frac{\varphi}2)}|1\ra$ and $|\phi_1\ra =
-\sin{(\frac{\varphi}2)}|0\ra + \cos{(\frac{\varphi}2)}|1\ra$. For
any parameters $\alpha\in [-1,1]$ and $\varphi\in [0,\pi]$ this is
an orthonormal basis in $\CC^4$, i.e.
$\la\tilde{\Psi}_i|\tilde{\Psi}_j\ra=\delta_{i,j}$. Consider the
density matrix
\be
\rho=\lambda_1 |\tilde{\Psi}_1\ra\la\tilde{\Psi}_1|
 + \lambda_2 |\tilde{\Psi}_4\ra\la\tilde{\Psi}_4| +
\lambda_3 |\tilde{\Psi}_3\ra\la\tilde{\Psi}_3| +
\lambda_4 |\tilde{\Psi}_2\ra\la\tilde{\Psi}_2|
\in D(\CC^4,\lambda).
\ee
The reduced density matrices $\rho^A=tr_B(\rho)$ and
$\rho^B=tr_A(\rho)$ have the lowest eigenvalues
\be
\lambda^A=\frac12
\left( 1- |\alpha|\cdot \sqrt{\lambda_{12}^2 + \lambda_{34}^2 +
2\lambda_{12}\lambda_{34}\cos{\varphi}} \right),
\ \
\lambda^B=\frac12
\left[ 1 - |\alpha|\cdot (\lambda_1 +
\lambda_2 - \lambda_3 - \lambda_4) \right],
\ee
where $\lambda_{12}=\lambda_1 - \lambda_2$ and
$\lambda_{34}=\lambda_3-\lambda_4$. One can easily verify that the
region of parameters $0\le \alpha \le 1$, $0\le \varphi \le \pi$ is
mapped to the triangle OBC on FIG.~1. Our goal is achieved.


\begin{thebibliography}{100}

\bibitem{Carteret} H. A. Carteret, A. Higuchi, and A. Sudbery,
{ ``Multipartite generalization of the Schmidt decomposition``},
LANL preprint archive  quant-ph/0006125.

\bibitem{Acin} A. Acin, A. Andrianov, L. Costa, E. Jane, J.I.~Latorre,
and R.~Tarrach,
{ ``Generalized Schmidt decomposition and classification
of three-quantum-bit states``},
LANL preprint archive quant-ph/0003050.

\bibitem{Werner}
K.G.H. Vollbrecht and  R.F.Werner, { ``Entanglement measures under symmetry``},
LANL preprint archive quant-ph/0010095

\bibitem{Sudbery00}
A.Sudbery,
{ ``On local invariants of pure three-qubit states``},
{\em J.Phys.A} {\bf 34}, 643-652 (2001),
also available at LANL preprint archive quant-ph/0001116

\bibitem{Barnum}
H.Barnum and  N.Linden, { ``Monotones and invariants for multi-particle quantum
states``}, LANL preprint archive quant-ph/0103155

\bibitem{Sudbery02}
A. Higuchi, A. Sudbery, and J. Szulc,
{ ``One-qubit reduced states of a pure many-qubit state: 
polygon inequalities``}, LANL preprint archive quant-ph/0209085



\bibitem{Werhl}
A.~Wehrl,
{ ``How chaotic is a state of a quantum system?''},
{\em Reports on Mathematical Physics}, vol. 6, pp. 15--28, 1974.


\bibitem{conv} Recall that for any subset $X$ of a linear space $\calL$,
the convex envelope $Conv(X)\subset \calL$ is a set of all linear combinations
$\sum_\alpha p_\alpha x_\alpha$, where $x_\alpha\in X$, $p_\alpha$ are
non-negative real numbers, and $\sum_\alpha p_\alpha =1$.
We will often say about convex sets of mean field states assuming that
their linear combinations are defined as
$t(\rho^{(1)}_1,\ldots,\rho^{(n)}_1) + (1-t)(\rho^{(1)}_0,\ldots,\rho^{(n)}_0)
\equiv (t\rho^{(1)}_1+ (1-t)\rho^{(1)}_0,\ldots,
t\rho^{(n)}_1+ (1-t)\rho^{(n)}_0)$.



\end{thebibliography}
\end{document}